\documentclass[aps,pra,twocolumn,showpacs,preprintnumbers,prb,superscriptaddress]{revtex4-2}
\usepackage{graphicx}  
\usepackage{dcolumn}   
\usepackage{bm}        
\usepackage{amssymb}   
\usepackage{framed}
\usepackage{amsmath}
\usepackage{hhline}
\usepackage{enumitem}
\usepackage{natbib}
\usepackage{subfigure,amsmath,verbatim,moreverb}
\usepackage{tabularx}
\usepackage{lipsum}
\usepackage{longtable}
\usepackage{booktabs}
\usepackage{threeparttable}
\usepackage{booktabs,amssymb,siunitx}
\usepackage{latexsym}
\usepackage{dcolumn}
\usepackage{amsmath}
\usepackage{epsf}
\usepackage{float}
\usepackage{hyperref}
\usepackage{xcolor}
\usepackage[utf8]{inputenc}
\usepackage[normalem]{ulem}
\usepackage{amsmath}
\usepackage{etoolbox}
\usepackage{soul}
\newcounter{subeqn} %

\usepackage{mathptmx}

\usepackage{booktabs}
\usepackage{tabularx}
\usepackage{adjustbox}
\usepackage{rotating}
\usepackage{color,soul}
\usepackage{multirow}
\usepackage{relsize}
\usepackage{ragged2e}

\usepackage{hyperref}
\hypersetup{
    unicode=true,          
    pdftoolbar=true,        
    pdfmenubar=true,        
    pdffitwindow=false,     
    pdfstartview={FitH},    
    colorlinks=true,       
    linkcolor=blue,          
    citecolor=blue,        
    urlcolor=blue,
}

\usepackage[us,12hr]{datetime} 

\begin{document}
\title{Strain and Correlation tuned magnetic anisotropy and Dzyaloshinskii–Moriya Interaction in 2D \textit{H}-FeTe$_2$}
\author{Dimple Rani}
\email{dimple.rani@niser.ac.in}
\affiliation{School of Physical Sciences, National Institute of Science Education and Research,
An OCC of Homi Bhabha National Institute, Jatni 752050, India}

\author{B. R. K. Nanda}
\email{nandab@iitm.ac.in} 
\affiliation{Condensed Matter Theory and Computational Lab, Department of Physics, Indian Institute of Technology Madras, Chennai 600036, India}

\author{Prasanjit Samal}
\affiliation{School of Physical Sciences, National Institute of Science Education and Research,
An OCC of Homi Bhabha National Institute, Jatni 752050, India}

 \begin{abstract}
In the ongoing research for two-dimensional (2D) ferromagnetic materials with strong intrinsic Dzyaloshinskii–Moriya interaction (DMI), most of the efforts are focused on doping, Janus engineering, or heterostructure formation to break inversion symmetry and enhance spin--orbit coupling (SOC). Here, we demonstrate that a pristine 2D material monolayer  \textit{H}-FeTe$_2$ can naturally host robust DMI and magnetic anisotropy due to its intrinsic broken inversion symmetry and the strong SOC of Te atoms. We explore the effect of biaxial strain and electron correlation on \textit{H}-FeTe$_2$ using  first principle DFT+$U$ calculations. We systematically investigate the Heisenberg exchange interaction, magnetic anisotropy, and DMI  in the space spanned by strain and correlation. Our results reveal a distinct, non-monotonic strain dependence of both magneto anisotropy energy (MAE) and DMI, including a strain-tunable crossover between in-plane and out-of-plane magnetic easy axes. A remarkable enhancement of the in-plane DMI is observed under the combined influence of strain and strong correlations, which is unusual for pristine 2D materials and suggests a favorable regime for spintronic applications. Notably, even in the absence of strain, \textit{H}-FeTe$_2$ exhibits finite DMI and considerable anisotropy rare for a pure 2D material. Through these findings, we present \textit{H}-FeTe$_2$ as a unique pristine 2D system with robust and tunable spin interactions for exploring fundamental spin orbit-driven magnetic phenomena.
\end{abstract}

\maketitle

\section{Introduction}
Spin–orbit coupling (SOC) is a fundamental mechanism driving various spin-dependent interactions in low-dimensional systems. In 2D materials, SOC leads to phenomena such as the Dzyaloshinskii–Moriya interaction (DMI), magnetic anisotropy energy (MAE), and Rashba spin splitting, all of which are crucial for controlling spin textures and magnetic anisotropies \cite{heide2008dzyaloshinskii, birss1964symmetry}. These effects become particularly prominent in systems with broken inversion symmetry and strong electronic correlations, offering opportunities to engineer spintronic functionalities directly within the atomic layers \cite{gong2017discovery, burch2018magnetism}.

Owing to recent studies, 2D transition metal dichalcogenides (TMDCs) of the form MX$_2$ have emerged as promising platforms for spintronic applications due to their intrinsic magnetic and electronic properties~\cite{PhysRevB.95.075434,voltage}. TMDCs consist of transition metals (M = Mn, Fe, Ni, V, Cr, Sc, Ti) and chalcogen atoms (X = S, Se, Te), forming layered structures held together by van der Waals forces. This structural feature allows these materials to be exfoliated into atomically thin monolayers, enabling quantum confinement effects and enhanced interfacial interactions. TMDCs exhibit a rich diversity of physical behaviors, ranging from semiconducting to metallic and superconducting phases, driven by their tunable band gaps and diverse crystal symmetries~\cite{Li2019}. Importantly, the presence of heavy transition metals results in strong SOC, a critical ingredient in determining magnetic anisotropy and DMI. TMDCs with magnetic elements such as Fe, Cr, and Mn show robust magnetism even at the monolayer level~\cite{geim2013van,manzeli20172d}. Furthermore, the DMI in these systems can be effectively tuned via external strain, magnetic fields, or interface engineering with high-SOC materials~\cite{moreau2016additive,fert2017magnetic}, making them attractive candidates for the development of next-generation spintronic technologies\cite{Fujisawa2020}.

The Dzyaloshinskii–Moriya interaction (DMI) arises in systems with strong SOC and broken inversion symmetry, often occurring at interfaces between magnetic and non-magnetic layers or in bulk non-centrosymmetric crystals ~\cite{dzyaloshinsky1958thermodynamic,dmi,dmi2}. This asymmetric magnetic exchange is expressed as $\sum_{\langle i, j \rangle} \mathbf{d}_{ij} \cdot (\mathbf{S}_i \times \mathbf{S}_j)$, where \(\mathbf{d}_{ij}\) represents the DMI vector between neighboring spins \(\mathbf{S}_i\) and \(\mathbf{S}_j\).  This interaction favors a canted or chiral alignment between neighboring spins rather than the parallel or antiparallel alignments typically driven by Heisenberg exchange interaction ~\cite{bogdanov1989thermodynamically}. In magnetic multilayer systems, DMI is commonly interfacial, arising from the interaction between magnetic layers (such as Co, Fe, or Ni) and heavy non-magnetic layers with high SOC (such as Pt, Ir, or W) ~\cite{moreau2016additive,8vvn-k9p3,7plx-jqlw}. The SOC of the non-magnetic layers induces a torque on the neighboring spins, resulting in a preferred rotational alignment, which stabilizes chiral magnetic textures like spin spirals and skyrmions ~\cite{thiaville2012dynamics,sampaio2013nucleation,10.1021/acsanm.5c05785}. DMI has gained considerable attention due to its crucial role in determining the chiral nature of magnetic structures. The interplay between DMI, perpendicular magnetic anisotropy (PMA), and conventional Heisenberg exchange energy governs the symmetry, stability, and orientation of magnetic textures~\cite{fert2017magnetic,PhysRevB.107.104428, doi:10.1021/acs.nanolett.2c03287,jana-gw,Rani_2024,10.1021/acs.jpcc.4c07037}. This competition can lead to the emergence of non-collinear spin configurations and influences spin dynamics in low-dimensional systems. As such, DMI remains a central aspect in the exploration of novel magnetic materials, particularly for applications in spintronics and magnetization control at the nanoscale~\cite{back20202020}.

Magnetic ordering has been extensively studied in a wide range of materials, including non-centrosymmetric B20 bulk compounds such as MnSi, FeGe, CoGe, and GdFeCo~\cite{yu2011near,yu2010real,mühlbauer2009skyrmion,kim2019bulk}. In addition, interfacial multilayer systems like Pt/Co/Ir, Pt/CoFe/MgO, Ta/CoFe/MgO, and Ir(111)/Fe have shown rich magnetic behavior driven by strong spin-orbit coupling and broken inversion symmetry~\cite{moreau2016additive,emori2013current,heinze2011spontaneous,PhysRevB.109.045133}. With the advancement of low-dimensional magnetism, recent research efforts have increasingly focused on two-dimensional (2D) magnets due to their intrinsic long-range magnetic ordering and tunable magnetic properties. Prominent 2D magnets include MnBi$_2$Te$_4$~\cite{Mn}, Fe$_n$GeTe$_2$ (with $n=3,4,5$)~\cite{Fe-based}, CrI$_3$~\cite{huang2017layer}, CrGeTe$_3$~\cite{gong2015crystallographic}, and VSe$_2$~\cite{vse2}. These systems serve an ideal platforms for exploring magneto-optical and magnetoelectric effects at the atomic scale, which are essential for next-generation ultra-compact spintronic devices. Moreover, Janus monolayers such as Cr$_2$X$_3$Y$_3$ (X, Y = Cl, Br, I; X $\neq$ Y)~\cite{jan}, MnXY (X, Y = S, Se, Te; X $\neq$ Y)~\cite{janus}, and CrXTe (X = S, Se)~\cite{crjan} provide asymmetric environments that enable novel magnetic interactions, enriching our understanding of magnetism in 2D materials.

\begin{table*}[ht]
\centering
\caption{ Optimized bond lengths (\textit{d}$_{\text{Fe-Fe}}$ and \textit{d}$_{\text{Fe-Te}}$), bond angles ($\theta_{1}$ and $\theta_{2}$), magnetic moment ($\mu$) of Fe atom, and Fe and Te resolved SOC contributions ($E_{\text{SOC}}^{\text{Fe}}$ and $E_{\text{SOC}}^{\text{Te}}$) to the energy in FeTe$_2$ monolayer under different strains.}
\begin{ruledtabular}
\begin{tabular}{lcccccccc}
\textbf{Strain ($\%$)}  & \textbf{\textit{d}$_{\text{Fe-Fe}}$ ($\AA$)} & \textbf{\textit{d}$_{\text{Fe-Te}}$ ($\AA$)} & \textbf{$\theta_{1}$ (deg)} & \textbf{$\theta_{2}$ (deg)} & \textbf{$\mu$ ($\mu_\text{B}$)} & \textbf{\textit{E}$_{\text{SOC}}^{\text{Fe}}$ (meV)} & \textbf{\textit{E}$_{\text{SOC}}^{\text{Te}}$ (meV)} \\ \hline
-6 & 3.27 & 2.44  & 78.43 & 84.28 & 2.020 & -9.15 & -209.81 \\ 
-4 & 3.34 & 2.47  & 77.25 & 85.15 & 2.030 & -9.27 & -211.03 \\ 
-2 & 3.41 & 2.50  & 76.10 & 85.99 & 2.046 & -9.30 & -213.22 \\ 
 0 & 3.48 & 2.53  & 74.98 & 86.81 & 2.067 & -9.38 & -215.07 \\ 
 2 & 3.55 & 2.56  & 73.88 & 87.60 & 2.071 & -9.66 & -215.55 \\ 
 4 & 3.62 & 2.59  & 72.83 & 88.37 & 2.081 & -10.16 & -215.60 \\ 
 6 & 3.69 & 2.63  & 71.78 & 89.11 & 2.089 & -10.42 & -215.64 \\ 
\end{tabular}
\end{ruledtabular}
\label{tab1}
\end{table*}

Based on previous studies ~\cite{PhysRevB.95.075434,voltage}, FeTe$_2$ shows great potential for spintronic applications, though there are hardly any report on the effect of DMI in 2D FeTe$_2$. To investigate the magnetic and DMI properties of \textit{H}-FeTe$_2$, we consider the H (point group D$_3$h) crystal structure of FeTe$_2$~\cite{jp212558p}. Our approach to analyze FeTe$_2$’s properties  by applying both compressive and tensile strain, ranging from -6\% to 6\%, where the stabilities of these strained structures have been verified by taking phonon dispersion into account, to explore how strain affects the DMI and magnetic anisotropy energy (MAE). We find that strain and correlation substantially influence the MAE. Interestingly, strain can make the MAE reverse its sign and thereby serve as a tool to switch the magnetic easy axis from in-plane to out-of-plane. It is noteworthy to highlight that through tensile strain one can achieve a giant change in MAE from -6 to 5.3 meV. The change is driven by the shifting of the spin-polarized states in energy. The competition between the dominant exchange couplings which include direct and superexchange interactions, and DMI makes the magnetic structure of FeTe$_2$ unique.
 Our findings can be extended to the layered chalcogenide compounds which are magentically active. Our analysis offers deep insights for advancing 2D materials in spintronic applications.

\section{ Structural $\&$ Computational Details}
\begin{figure*}
    \centering
    \includegraphics[width=1\linewidth]{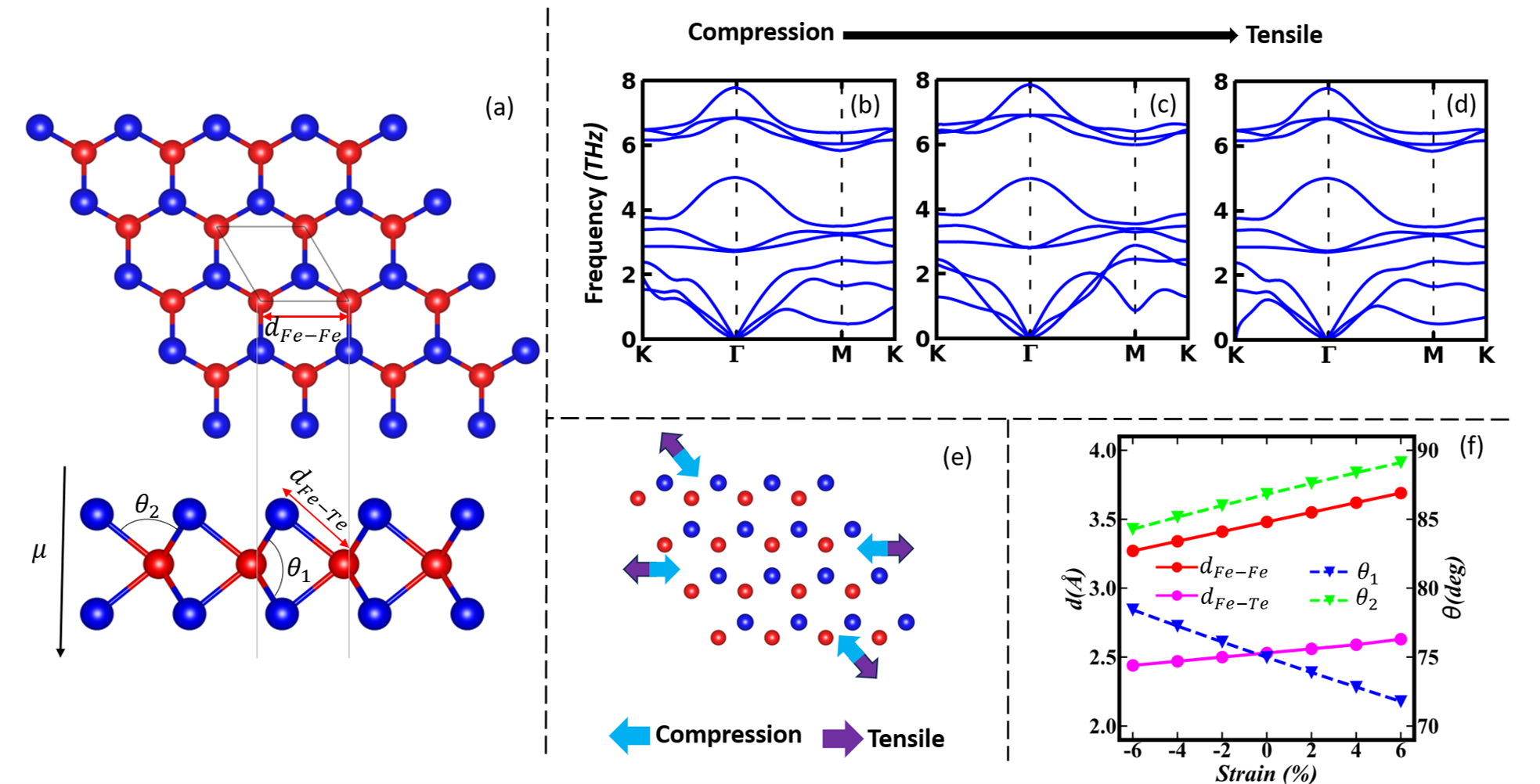}
    
\caption{\justifying \textbf{Structure of FeTe$_2$ monolayer under strain.} (a) Top and side views of the 2H phase of the FeTe$_2$ monolayer. In the top view, the atomic arrangement is shown with Fe atoms (red spheres) and Te atoms (blue spheres), where $d_{\text{Fe--Fe}}$ denotes the in-plane lattice constant. The side view illustrates the bond lengths $d_{\text{Fe--Te}}$ and the bond angles $\theta_1$ and $\theta_2$.  
(b)-(d) Phonon band structure of the FeTe$_2$ monolayer at -6, 0 and 6 $\%$ strain, confirming its dynamic stability.  
(e) Schematic diagram showing the application of biaxial strain on the top surface of the FeTe$_2$ monolayer.  
(f) Variation of bond lengths ($d_{\text{Fe--Te}}$) and bond angles ($\theta_1$, $\theta_2$) as a function of applied biaxial strain (\%).}

\label{fig1}
\end{figure*}

Fig.~\ref{fig1}(a) shows the top and side views of monolayer \textit{H}, which has a hexagonal crystal structure with point group D${3h}$. In this structure, a layer of Fe atoms is sandwiched by two layers of Te atoms. The Fe atoms are surrounded by Te atoms in a trigonal prismatic shape, which is common in many of the transition metal dichalcogenides (TMDs). This symmetric arrangement affects how the Fe $d$-orbitals split, which in turn influences the electronic and magnetic behavior of the material. Since the structure lacks inversion symmetry and includes heavy Te atoms, it provides a suitable platform to study the SOC influenced effects such as magnetic anisotropies and DMI~\cite{voltage}.

Our first-principles calculations on 2D FeTe$_2$ are conducted using Density Functional Theory (DFT) as implemented in the Vienna Ab initio Simulation Package (VASP)~\cite{kresse1996efficiency}. For electron-core interactions, we employ the projected augmented-wave (PAW) method~\cite{kresse1996efficient,kresse1999ultrasoft}, and the exchange-correlation energy is treated with the Perdew-Burke-Ernzerhof (PBE) formulation within the generalized gradient approximation (GGA)~\cite{perdew1996generalized}. In order to describe the strongly correlated 3d
electrons of Fe, the DFT+U method is applied~\cite{voltage,fe-mag}. The Kohn-Sham orbitals are represented with a plane-wave basis set with a kinetic energy cutoff of 480 eV, while the k-point sampling is performed on a $\Gamma$-centered 18 × 18 × 1 Monkhorst-Pack grid.  To effectively study the magnetic exchange interactions \( J_1 \), \( J_2 \), and \( J_3 \) in FeTe$_2$, we utilize a 4$\times$4 supercell. This size is crucial because it captures the necessary first, second, and third nearest-neighbor interactions that are vital for our analysis of Heisenberg exchange. Furthermore, when examining the DMI, we can rotate the spin orientation of the Fe atoms in both clockwise and anticlockwise directions for 4$\times$1 supercell. This flexibility allows us to explore the chiral asymmetry in the system, which is essential for accurately determining the DMI vector and understanding its impact on the magnetic properties. Phonon dispersions are calculated using PHONOPY code~\cite{TOGO20151}.

To determine the  DMI vectors ($d_{ij}$), we perform a series of calculations in three main steps. First, structural relaxations are carried out to obtain the most stable interfacial geometries, stopping once the energy and forces reach thresholds below 10$^{-6}$ eV and 0.001 eV/$\text{\AA}$, respectively. Next, we solve the Kohn-Sham equations without SOC to determine the ground-state charge density of the system. In the final step, SOC is included, and we compute the self-consistent total energy as a function of the magnetic moment orientation, using VASP’s constrained method to control the orientation. This method, previously used to evaluate DMI in bulk frustrated systems and insulating chiral-lattice magnets~\cite{PhysRevB.84.224429}, is adapted here to analyze the DMI of the monolayer.
 
\section{\label{sec:level2}Result and Discussion}
\subsection{Structural Stability $\&$ Strain}
Biaxial strain ($\epsilon$) is introduced by uniformly varying the in-plane lattice constants ($a = b$) from $-6\%$ to $+6\%$ in increments of 2\%, as illustrated in Fig.~\ref{fig1}(e). To assess the dynamical stability of both the unstrained and strained structures, phonon spectra were computed using a $4 \times$ 4 supercell, as shown in Figs.~\ref{fig1}(b)-\ref{fig1}(d). The absence of significant imaginary frequencies across the Brillouin zone confirms the dynamical stability of \textit{H}-FeTe$_2$ monolayer within the entire considered strain range. Beyond 6 $\%$ strain, we found the apperance of imaginary modes.

The lack of inversion symmetry in both in-plane and out-of-plane directions of monolayer \textit{H}-FeTe$_2$ results in an asymmetric bonding environment around the Fe atoms. Although all Fe--Te bond lengths are equal, the bond angles around Fe differ ($\theta_1 \neq \theta_2$), which breaks the local inversion symmetry at the Fe site. This structural asymmetry generates an asymmetric crystal field and modifies the local electronic environment. According to Moriya’s theory~\cite{dmi2}, such a local inversion symmetry breaking, in the presence of SOC, allows the DMI to arise. For systems with hexagonal symmetry, the DMI vector lies in the plane and is oriented perpendicular to the spin direction. The strain serves as a natural parameter to tailor the magnetic properties in general and the DMI in particular. 

Changes in structural and magnetic properties as a function of strain are listed in Table~\ref{tab1}.  The Fe-Te bond length, $d_{\text{Fe-Te}}$, increases with tensile strain, while the bond angles $\theta_1$ and $\theta_2$  change differently: $\theta_1$ decreases, and $\theta_2$ increases, as illustrated in Fig.~\ref{fig1}(f). The magnetic moment of Fe ($\mu$) increases with  tensile strain as can be seen from the Table~\ref{tab1}. Notably the Te atoms dominate the SOC contribution, with $E_{\text{SOC}}^{\text{Te}}$ being significantly higher than $E_{\text{SOC}}^{\text{Fe}}$.\\
  \begin{figure}
    \centering
    \includegraphics[width=1\linewidth]{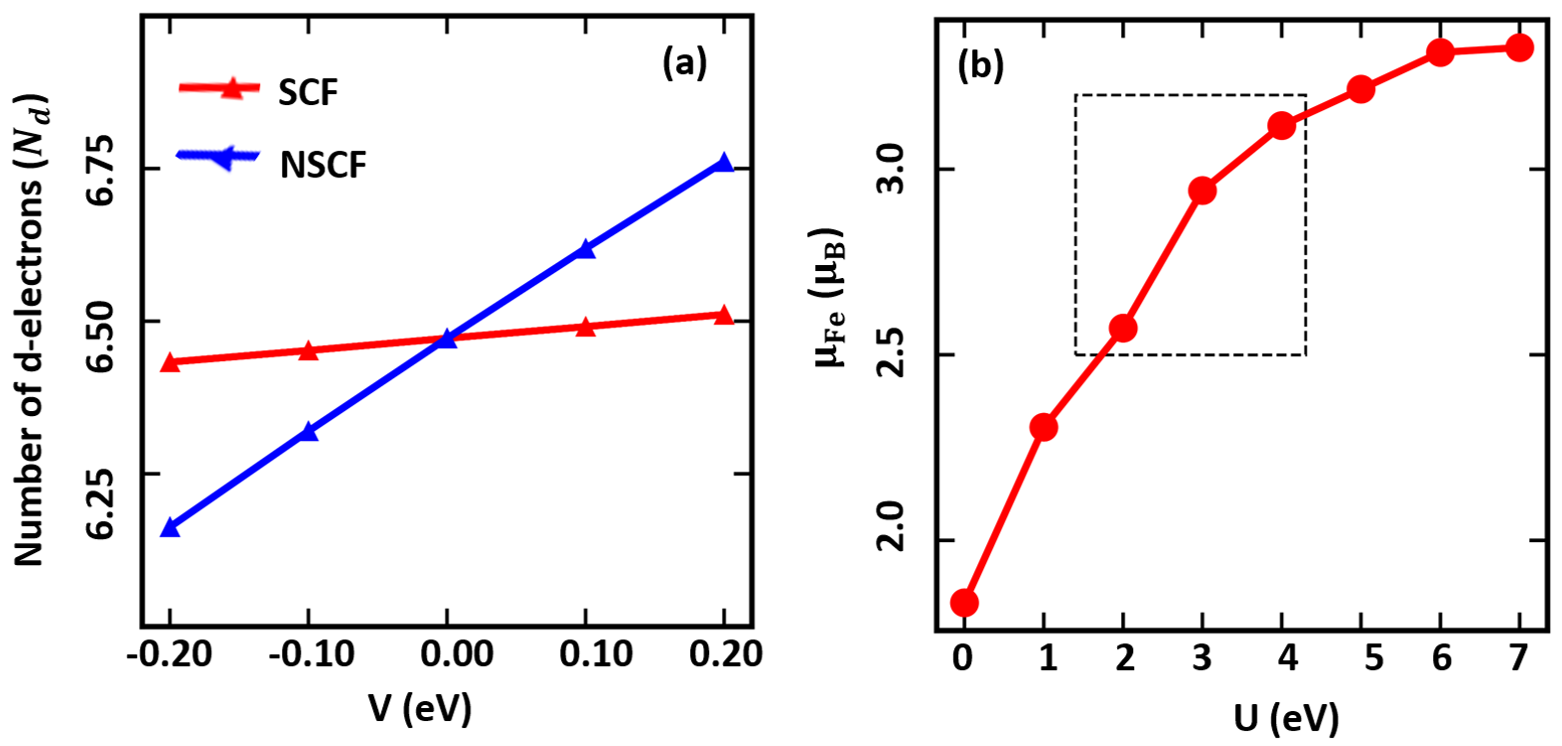}
   \caption{\justifying(a) Computation of the on-site Hubbard \(U\) parameter using linear response theory: the number of Fe \(d\)-electrons (\(N_d\)) is plotted as a function of the applied on-site potential shift \(V\), using both self-consistent (SCF) and non-self-consistent (NSCF) approaches. The difference in the slopes of these lines yields the \(U\) value.  (b) Variation of the magnetic moment of Fe atoms ($\mu_{\text{Fe}}$) as a function of the Hubbard $U$ parameter. The dotted box highlights the region where the magnetic moment exhibits the maximum change with increasing $U$.}
 \label{mu-fig}
\end{figure}
 \begin{figure*}
    \centering
    \includegraphics[width=1.0 \linewidth]{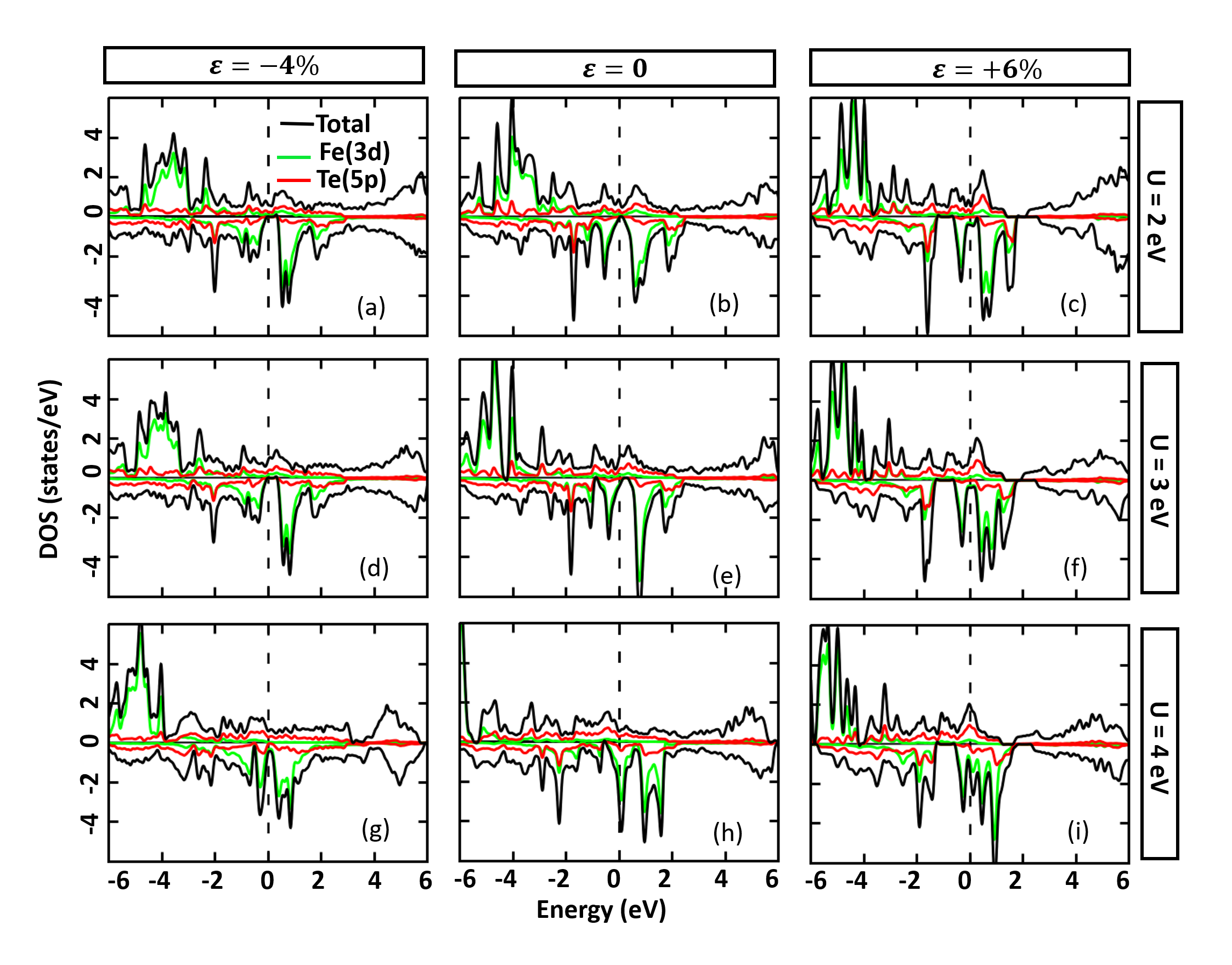}
    \caption{%
        \justifying
        \textbf{Evolution of electronic structure of FeTe$_2$ as a function of Strain and correlation.} Total and orbital projected resolved density of states of the FeTe\textsubscript{2} monolayer under three different strain conditions: $\varepsilon = -4\%$, $0\%$, and $+6\%$ (left to right columns), and for varying U (2 - 4 eV). The Fermi level ($E_F$) is set to 0.
    }
        \label{dos}
\end{figure*}

\subsection{Effect of Electron Correlation}
Since FeTe$_2$ has localized $d$-electrons, it is expected that the onsite Coulomb repulsion will also have a deterministic role in the electronic and magnetic structure of this compound. For this purpose, we have employed the PBE\,+\,\textit{U} approach~\cite{+u}. 
We calculated the effective Hubbard parameter $U$ for \textit{H}-FeTe$_2$ using the linear response theory approach ~\cite{PhysRevB.71.035105}. The computed value of $U = 3.20$~eV is obtained self-consistently, as illustrated in Fig.~\ref{mu-fig}(a). In this method, the parameter $U$ is extracted from the difference between the inverse of the non-interacting and interacting response functions:
\begin{equation}
U = \chi_0^{-1} - \chi^{-1} = \left( \frac{\partial n_i^{\mathrm{Scf}}}{\partial \alpha_i} \right)^{-1} - \left( \frac{\partial n_i^{\mathrm{Nscf}}}{\partial \alpha_i} \right)^{-1},
\end{equation}
where $\alpha_i$ is a small perturbation in the on-site potential at atomic site $i$, and $n_i^{\mathrm{Scf}}$ and $n_i^{\mathrm{Nscf}}$ are the corresponding the occupation numbers obtained from SCF and NSCF field calculations, respectively.

 In previous studies~\cite{voltage,fe-mag}, a Hubbard \textit{U} value of 2\,eV has commonly been adopted. To systematically investigate the impact of electronic correlations on the magnetic properties of monolayer FeTe$_2$, we computed the magnetic moment of Fe ($\mu_{\text{Fe}}$) as a function of on-site Coulomb interaction parameter $U$, ranging from 0 to 7\,eV. As shown in Fig.~\ref{mu-fig} (b), $\mu_{\text{Fe}}$ increases monotonically with $U$, with a rapid enhancement observed between $U = 2$ and 4\,eV. Beyond $U = 4$\,eV, the magnetic moment continues to increase but with a significantly reduced slope, indicating a gradual approach toward a high-spin limit. This intermediate regime (highlighted in the figure) represents a crossover region where electronic correlations begin to strongly influence the local magnetic moment. Based on this analysis, we focus on the $U$ range of 2 - 4\,eV to explore  the effect of correlation  on magnetism in FeTe$_2$.
\begin{table*}
\caption{\label{tab-all-4} Magnetic Exchange interactions (\textit{J}$_{\text{1}}$, \textit{J}$_{\text{2}}$ and \textit{J}$_{\text{3}}$), single-ion anisotropy (K$_u$) for U = 2 -- 4 eV as a function of strain (-6 $\%$ to 6 $\%$.). The values are expressed in units of meV. } 
\begin{ruledtabular}
\begin{tabular}{cccccccccccccccc}
& \multicolumn{4}{c}{U = 2 eV} & \multicolumn{4}{c}{U = 3 eV} & \multicolumn{4}{c}{U = 4 eV} \\ \cline{2-5} \cline{6-9} \cline{10-13}
Strain ($\%$)& J$_1$ & J$_2$ & J$_3$ & K$_u$ &  J$_1$ & J$_2$ & J$_3$ & K$_u$ &  J$_1$ & J$_2$ & J$_3$ & K$_u$ \\ \hline
{-6} & -7.87 & 5.13 & -8.71 & -3.01 &-7.67 & 6.12 & -6.00 & -1.68 & -5.44 & 5.85 & -3.93 & -1.32 &   \\
{-4} & -9.02 & 4.09 & -8.56 & -4.89 & -8.88 & 5.70 & -5.19 & -1.99 & -6.63 & 4.22 & -3.41 & -1.54 &  \\
{-2} & -11.88 & 3.28 & -4.77 & -6.47 & -6.92& 4.03 & -3.40 & -2.73 & -4.58 & 3.93& -2.39 & -4.68& \\
{0} & -9.40 & 3.33 & -4.02 & -6.54 & -5.05 & 3.17 & -2.84 & -4.8 & -4.21 &2.96 & -2.16 & 2.68&  \\
{2} & -7.51 & 3.15 & -3.37 & 1.18 & -4.40 & -1.88 & -2.55 & 3.38& -3.45 &-2.40& -1.43 & 4.72 &  \\
{4} & -5.81 & 2.64 & -3.28 & 2.89 & -3.47 & -2.74 & -1.80& 4.17 & -2.96 &-3.13 & -1.18 & 4.91 &  \\
{6} & -2.49 & 2.06 & -2.34 & 4.50 & -1.49 & -2.78 & -1.03 & 4.98 & -1.59 &-3.89 & -0.96 & 5.30 & \\
\end{tabular}
\end{ruledtabular}
\label{tab-all-4}
\end{table*}
\subsection{Electronic Strurcture}
To understand how electronic correlation and lattice strain jointly influence the magnetic and electronic behavior of the FeTe$_2$ monolayer, we examine the spin resolved orbital projected density of states  as shown in Fig.~\ref{dos}. The rows represent increasing onsite Coulomb interaction $U =$ 2 - 4 \,eV, while the columns span biaxial strain from $\varepsilon = -4\%$ , 0$\%$ $+6\%$. Across all cases, the system remains metallic in the spin up channel. However, in the spin down channel, for weaker correlation and strong compressive strain a finite gap is observed at E$_F$. As the strain becomes tensile and with increase in U, the Fe-d spin down states lie on the E$_F$ to make the system metallic in both the spin channels. In addition, as expected with strong correlation, the Fe-d states in the spin up channel are pushed below in Energy. The strain  and electron correlation also brings minor, but significant changes to the Te-p states. As shown in Fig. ~\ref{dos-suppl}, the population of p$_z$ states vary strongly as compared to the doubly degenerate p$_x$ and p$_y$ states.

Collectively, the evolution of the electronic structure points to a strain induced crossover from a more itinerant electronic structure under compression to a more localized one under tension. Moreover, the increasing spectral weight asymmetry between spin up and spin down channels with higher \(U\) and tensile strain suggests an emerging imbalance in spin occupation, indicating that electronic correlations play a significant role in reshaping the spin-dependent electronic structure and thus the magnetic interaction strengths which include Heisenberg exchange interactions J, DMI (d) and the spin-anisotropy (k) as will presented in the following sections. Further, we will later show that the varying electronic structure significantly influences MAE in Section ~\ref{magnetic}.

\subsection{Magnetic Interactions}
The magnetic system with interacting spins can be represented through the following Hamiltonian.
\begin{equation}
H = \frac{1}{2}\sum_{i,j} J_{ij} \mathbf{S}_i \cdot \mathbf{S}_j 
    + K_{u} \sum_i \left(S_i^z\right)^2  
    + \frac{1}{2} \sum_{i,j} \mathbf{d}_{ij} \left(\mathbf{S}_i \times \mathbf{S}_j\right) 
    + \mu \nu B \sum_i S_i^z.
    \label{Hamiltomnian}
\end{equation}
Here, $J_{ij}$ is the Heisenberg exchange interaction parameter between spins on sites i and j.
The second term describes the spin anisotropy with \( K_u \) as the single-ion anisotropy constant. The term \( K_u \sum_i (S_i^z)^2 \)  signifies deviations of the spin along the \( z \)-axis. It favors alignment of the spin vector out of  the plane (if \( K_u > 0 \)) or along the \( x \)-axis (if \( K_u < 0 \)). In the third term of the Hamiltonian,  \( \mathbf{d}_{ij} \) is the Dzyaloshinskii-Moriya interaction (DMI) vector between spins at sites \( i \) and \( j \). Last term  represents the Zeeman energy, which favors alignment of the spins along the direction of the applied magnetic field $B$ where \(\mu\) and \(\nu\) are the  magnetic moment associated with each spin and number density respectively. 
\begin{figure*}
    \centering
    \includegraphics[width=1.0 \linewidth]{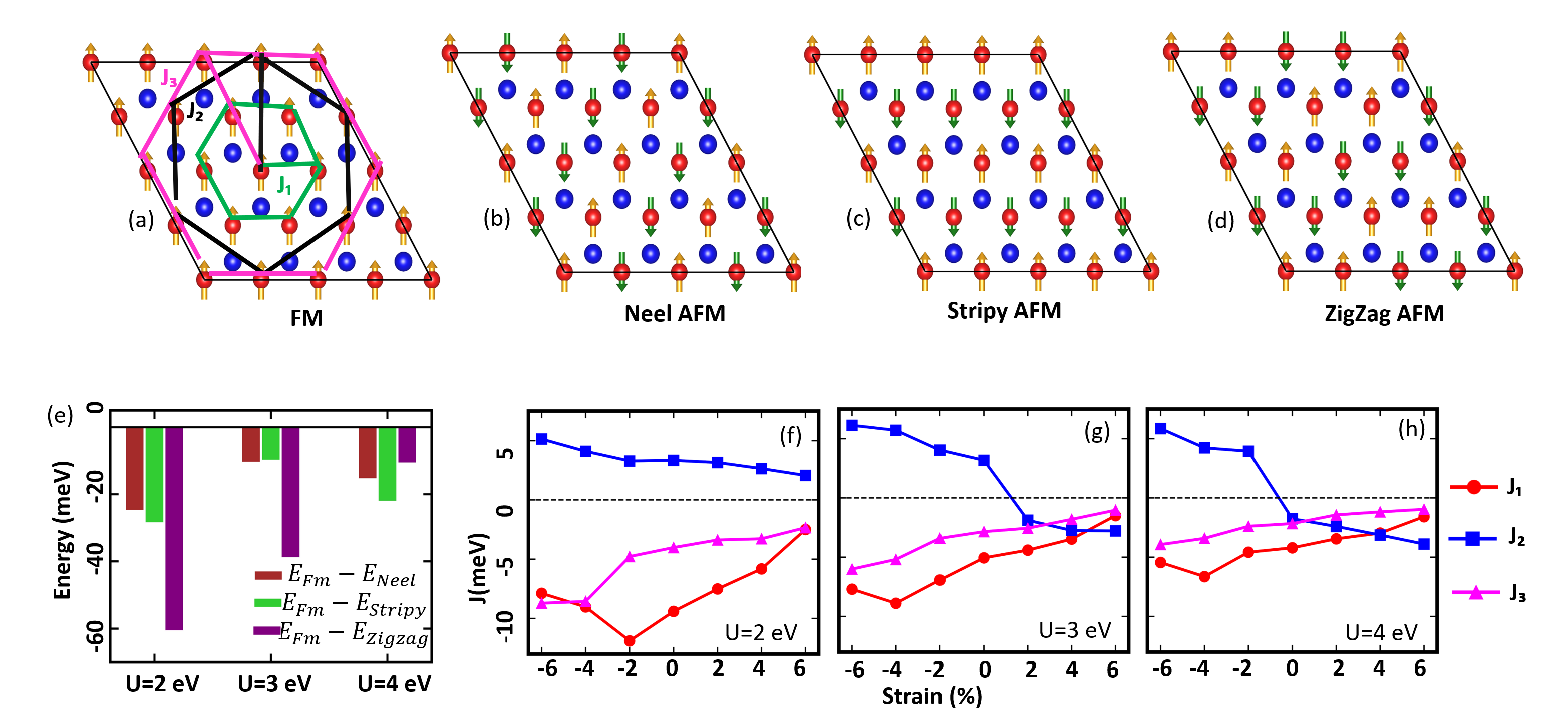}
    \caption{\justifying \textbf{Strain and correlation tunable magnetic exchange interactions.} Top view of the FeTe$_2$ monolayer:  
(a) FM spin configuration along with a schematic representation of the lattice showing Heisenberg exchange interactions $J_1$, $J_2$, and $J_3$;  
(b)--(d) Different types of AFM spin configurations: Néel-type, stripy, and zigzag patterns, respectively;  
(e) Energy differences between the four spin configurations (FM, Néel, stripy, and zigzag) for unstrained FeTe$_2$ as a function of U;  
(f)--(h) Variation of Heisenberg exchange interactions $J_1$, $J_2$, and $J_3$ as a function of strain (\%) for $U = 2$, 3, and 4 eV, respectively.}
 \label{fig3}
\end{figure*}
  To estimate the value of $J_{ij}$ from the DFT calculations, we have employed Noodelman's broken symmetry dimer method ~\cite{Noodleman1981} on our system where the difference between high spin (HS) and broken symmetry (BS) configuration is given by
\begin{equation}
    E_{HS} -E_{BS} = \frac{1}{2} S_{max}^{2} J
\end{equation}
In this study we are estimating the $J_{ij}$ up to third neighbor interactions (i.e. $J_1$, $J_2$, and $J_3$). Therefore, for the DFT calculations, we have considered four magnetic configurations, namely,  ferromagnetic (FM) , Néel type antiferromagnetic (AFM), Stripy AFM, and Zigzag AFM (see Fig. ~\ref{fig3}), and mapped the DFT obtained total energy for each of these configurations with equation (3) which resulted in four linear equations as follows:
\begin{subequations}
\begin{align}
E_{\text{FM}} &= S^2 \left( 6J_{1} + 6J_{2} + 6J_{3} \right) + E_0,
\label{fm} \\[6pt]
E_{\text{N\'{e}el}} &= S^2 \left( -2J_{1} + 2J_{2} - 2J_{3} \right) + E_0,
\label{gafm} \\[6pt]
E_{\text{Stripy}} &= S^2 \left( -2J_{1} - 2J_{2} + 6J_{3} \right) + E_0,
\label{aafm} \\[6pt]
E_{\text{Zigzag}} &= S^2 \left( 2J_{1} - 2J_{2} - 2J_{3} \right) + E_0.
\label{eafm}
\end{align}
\end{subequations}

Here, S is the integral spin associated to the magnetic atoms and E$_0$ is the non-magnetic energy.
The magnetic exchange parameters \( J_1 \), \( J_2 \), and \( J_3 \), obtained by solving the above set of linear equations, are presented in Fig.~\ref{fig3}(f)--(h) as a function of applied strain and the electron correlation parameter \( U \), with the corresponding numerical values summarized in Table~\ref{tab1}. As evident from the Fig.s, the exchange interactions exhibit a non-monotonic dependence on strain.  
 Under compressive strain (\( \varepsilon < 0 \)), \( |J_1| \) and \( |J_3| \) are relatively large, indicating strong FM interactions, particularly at \( \varepsilon = -2\% \). As the strain becomes tensile, \( J_1 \) becomes less negative and even approaches zero, while \( J_2 \) changes sign and becomes increasingly FM like for \(U > 2\) eV. The strong correlation weakens the strength of $J_1$ and $J_3$.
The trends of Js as a function of strain reflect a strain-driven crossover in magnetic interaction pathways, likely arising from the evolving d$_{Fe-Fe}$, d$_{Fe-Te}$  bond lengths and bond angles (\( \theta_1, \theta_2 \)), which modulate both direct and superexchange contributions. To evaluate the dominant magnetic ground state, we present a bar graph showing the energy differences of all AFM configurations relative to the FM configuration for the unstrained system, for U=2, 3, and 4 eV. As shown in Fig.~\ref{fig3}(e), the FM configuration consistently exhibits lower energy than all AFM configurations across the considered 
U values, indicating FM as the ground state magnetic configuration.

The combined influence of strain and electronic correlation provides an effective means of tuning magnetic interactions in FeTe$_2$ monolayers. Under compressive strain and lower \( U \) values, the system enters a regime of magnetic frustration due to competing exchange interactions, which may favor noncollinear or complex magnetic configurations. In contrast, at higher \( U \) and tensile strain, the exchange pathways become more ferromagnetic in character, promoting the stabilization of a robust magnetic order. These findings highlight how structural deformation and electronic correlation together enable access to a wide range of magnetic phases in two-dimensional materials.

\subsection{Magnetic Anisotropy Energy \label{magnetic} }
\begin{figure}
     \centering
     \includegraphics[width=0.8\linewidth]{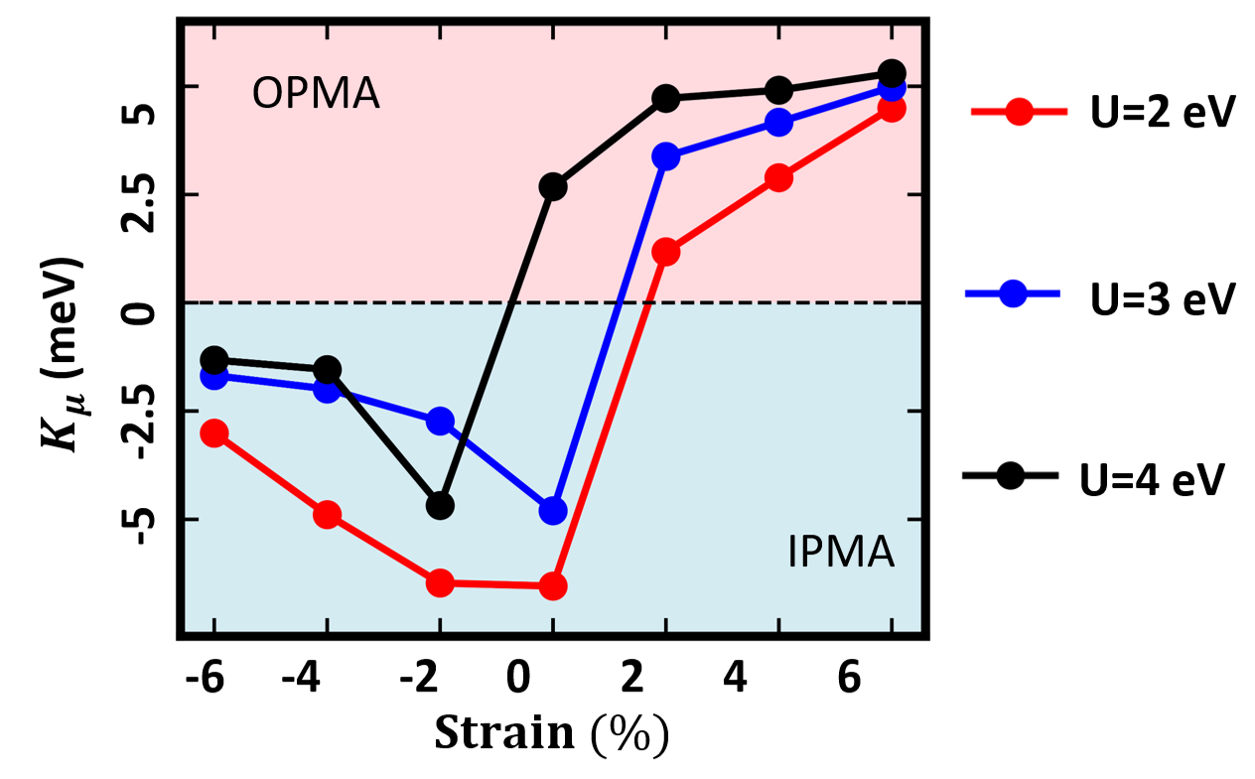}
     \caption{\justifying Single-ion magnetic anisotropy energy K$_{\mu}$ as a function of
biaxial strain for different U values. The pink and blue shaded
regions denote out-of-plane magnetic anisotropy  and in-plane 
magnetic anisotropy, respectively. }
     \label{ku-seper}
 \end{figure}
The second term in the Hamiltonian of equation~\ref{Hamiltomnian} corresponds to the single-ion anisotropy. In FeTe$_2$ monolayers, the presence of strong SOC —primarily originating from the heavy Te atoms — introduces significant spin-dependent energy shifts, which in turn enhance the magnetic anisotropy. This anisotropy is further modulated by strain and electronic correlation, enabling control over the magnetic easy axis. A large and tunable MAE is essential for stabilizing the magnetic order against thermal fluctuations and for supporting anisotropic spin textures in 2D systems.
The Magnetic single-ion anisotropy parameter, \(K_u\), can be calculated as the energy difference between the in-plane (E$^x$) and out-of-plane (E$^z$) magnetization orientations, given by:
\[
K_u \approx E_{[100]} - E_{[001]} = E^x - E^z
\]
Where \(K_u > 0\) indicates a perpendicular easy axis (out-of-plane magnetization), and \(K_u < 0\) corresponds to an easy-plane configuration (in-plane)
The strain-dependent variation of  $K_u$ for different $U$ values (2 - 4 eV), shown in Table ~\ref{tab1}, offers deep insight into the interplay between lattice distortion, electronic correlation, and SOC in determining magnetic anisotropy in two-dimensional systems.

At compressive strain ($-6\%$ to $-2\%$), $K_u$ remains negative across all $U$ values, indicating a preference for in-plane magnetic anisotropy (IPMA). 
 As strain increases toward the tensile side, $K_u$ transitions from negative to positive, signaling a crossover from IPMA to out-of-plane magnetic anisotropy (OPMA). This transition is not only strain-driven but also strongly influenced by increasing $U$, which models onsite Coulomb repulsion and reflects stronger electron localization. A larger $U$ enhances spin polarization and local moment formation on magnetic atoms, which in turn strengthens SOC-induced anisotropy, especially when mediated via hybridization with heavy atom Te. The variation of \(K_u\) with applied strain reveals some important trends.  For U = 2 eV, in the unstrained FeTe$_2$ monolayer, the system exhibits an easy-plane configuration with \(K_u\) = -6.54 meV, indicating a strong in-plane anisotropy which is consistent with the earlier studies ~\cite{voltage}. 
At $\epsilon = 2\%$, $K_u$ rises significantly with $U$, from 1.18 meV at $U = 2$ eV to 4.72 meV at $U = 4$ eV, and this enhancement continues under tensile strain, reaching 5.30 meV at 6\% strain for $U = 4$ eV. The  variation of k$_u$ from IPMA to OPMA  with tensile strain for different values of is given in Fig. ~\ref{ku-seper} . The tensile strain modifies the bond angles and crystal fields, altering orbital overlaps and thereby enhancing the anisotropic SOC contributions—particularly between the $p$ orbitals of the Te atoms and the $d$ orbitals of the Fe centers. This strain-driven orbital hybridization alters the electronic structure such that SOC energetically favors spin alignment perpendicular to the plane. The resulting out-of-plane anisotropy plays a critical role in stabilizing magnetic order in 2D systems, influencing both the thermal stability and the nature of spin excitations.
\begin{figure}
    \centering
    \includegraphics[width=0.7\linewidth]{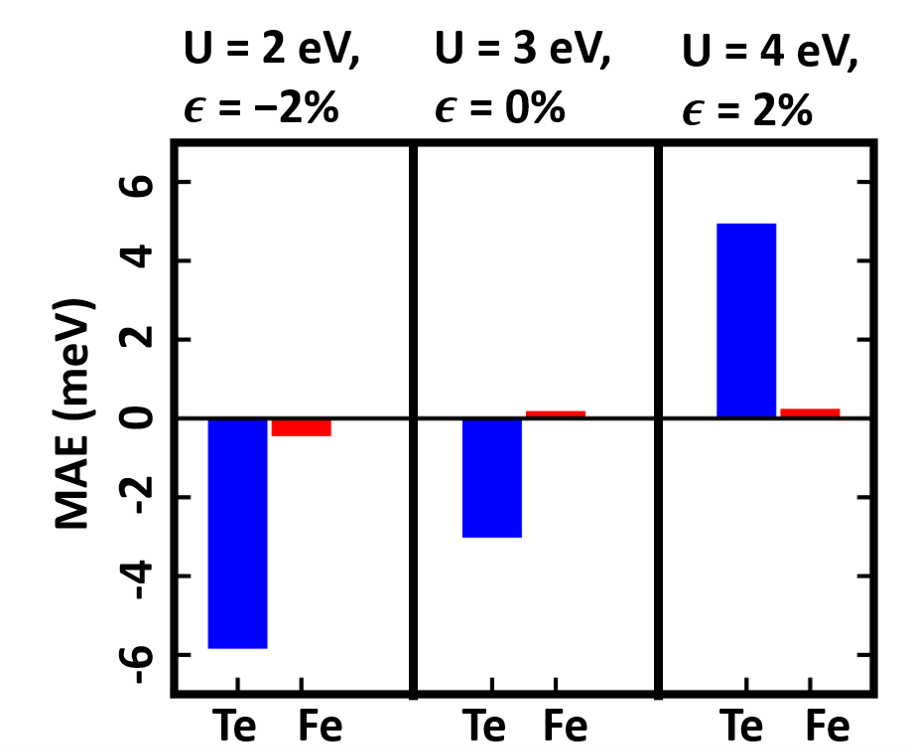}
    \caption{\justifying Atom resolved MAE for FeTe$_2$ under varying on-site Coulomb interaction \( U \) and biaxial strain \( \varepsilon \). The three panels correspond to \( U = 2\,\mathrm{eV},\ \varepsilon = -2\% \); \( U = 3\,\mathrm{eV},\ \varepsilon = 0\% \); and \( U = 4\,\mathrm{eV},\ \varepsilon = 2\% \), respectively. 
    }
    \label{atom-mae}
\end{figure}
\begin{figure*}
    \centering
    \includegraphics[width=1.05\linewidth]{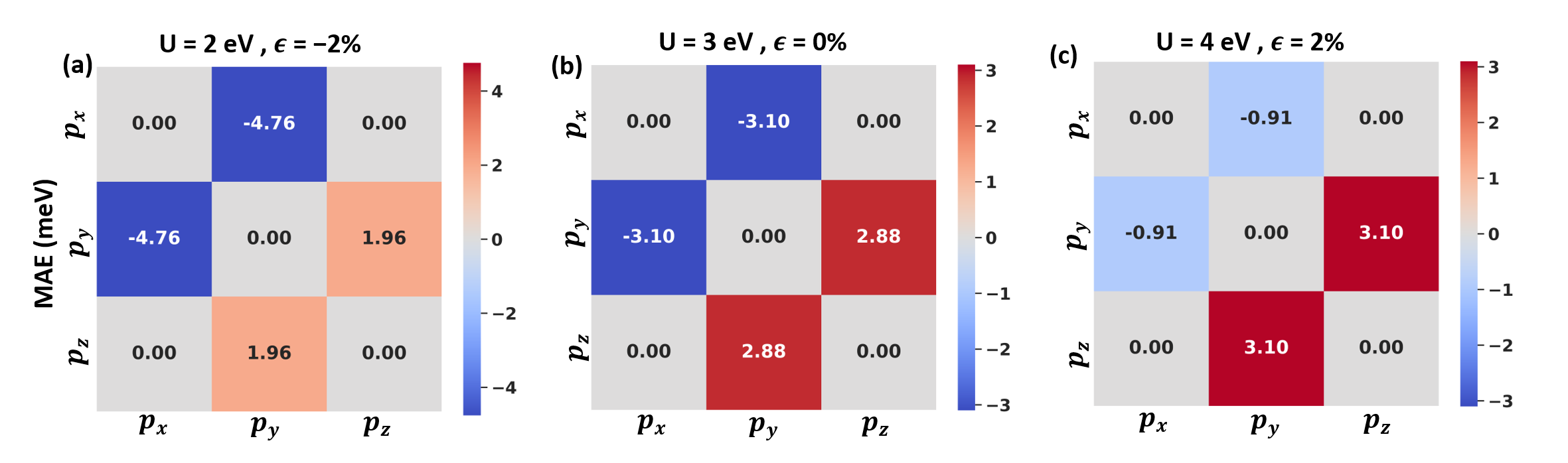}
  \caption{\justifying MAE contribution from 5p orbital hybridization in of Te atoms in the FeTe$_2$ monolayer under different conditions: (a) $U=2$ eV, $\epsilon = -2\%$ (compressive strain), (b) $U=3$ eV, $\epsilon = 0\%$ (unstrained), and (c) $U=4$ eV, $\epsilon = +2\%$ (tensile strain).}
    \label{orb}
\end{figure*}
\begin{figure*}
    \centering
\includegraphics[width=1.0\linewidth]{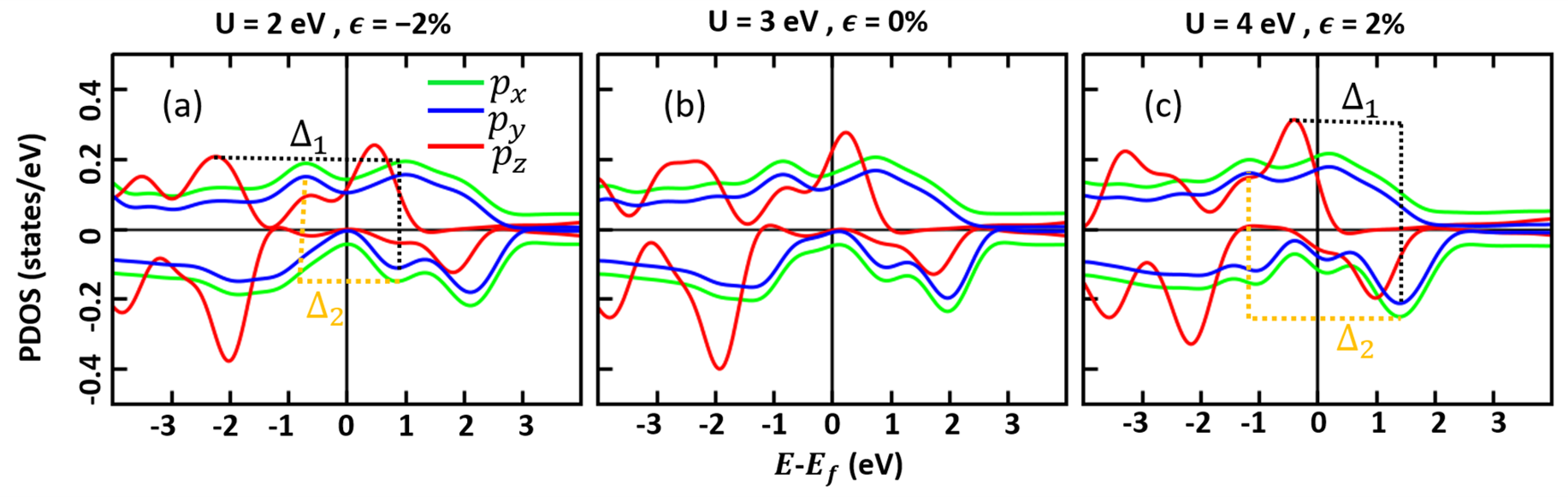}
    \caption{Orbital resolved projected density of states (PDOS) for the Te $p$ orbitals ($p_x$: green, $p_y$: blue, $p_z$: red) in the FeTe$_2$ monolayer under different conditions: (a) $U = 2$~eV, $\varepsilon = -2\%$ (compressive strain), (b) $U = 3$~eV, $\varepsilon = 0\%$ (unstrained), and (c) $U = 4$~eV, $\varepsilon = +2\%$ (tensile strain). The Fermi level is set at $E - E_F = 0$~eV and marked by the vertical black line.}
    \label{dos-suppl}
\end{figure*}

Fig.~\ref{atom-mae} shows the atom resolved MAE for Fe and Te atoms under varying strain and Hubbard \( U \). Te atoms dominate the MAE in all cases, highlighting the key role of Te-driven SOC. At \( U = 2\,\text{eV}, \varepsilon = -2\% \), Te contributes strongly to in-plane anisotropy (negative K$_u$). As \( U \) and strain increase, the MAE shifts, and at \( U = 4\,\text{eV}, \varepsilon = 2\% \), Te contributes positively, indicating a transition to out-of-plane anisotropy. Fe contributions remain negligible. These results demonstrate tunable magnetic anisotropy via strain and correlations. 

  To gain microscopic insight into the origin of the anisotropy behavior of the Te atom, we carry out a comparative analysis of the orbital-resolved hybridization contributions to the MAE. Within the framework of second-order perturbation theory proposed by Wang \textit{et al.}~\cite{PhysRevB.47.14932}, the MAE can be expressed in terms of angular momentum matrix elements as a sum of two contributions:  

\begin{subequations}
\begin{align}
\Delta E^{\downarrow \downarrow} &= E^{\downarrow \downarrow}(x) - E^{\downarrow\downarrow}(z) \nonumber \\
&= \lambda^2 \sum_{o^{\downarrow},u^{\downarrow}} 
\frac{ \left| \langle o^{\downarrow} | L_z | u^{\downarrow} \rangle \right|^2 
    - \left| \langle o^{\downarrow} | L_x | u^{\downarrow} \rangle \right|^2 }
{\varepsilon_{u^{\downarrow}} - \varepsilon_{o^{\downarrow}}},
\label{oma} \\[6pt]
\Delta E^{\uparrow \downarrow} &= E^{\uparrow \downarrow}(x) - E^{\uparrow \downarrow}(z) \nonumber \\
&= - \lambda^2 \sum_{o^{\uparrow},u^{\downarrow}} 
\frac{ \left| \langle o^{\uparrow} | L_z | u^{\downarrow} \rangle \right|^2 
- \left| \langle o^{\uparrow} | L_x | u^{\downarrow} \rangle \right|^2 }
{\varepsilon_{u^{\downarrow}}- \varepsilon_{o^{\uparrow}}},
\label{oma2} \\[6pt]
\mathrm{MAE} &= \Delta E^{\downarrow \downarrow} + \Delta E^{\uparrow \downarrow}.
\label{mae_total}
\end{align}
\end{subequations}
where $\lambda$ denotes the SOC strength, $\uparrow$ and $\downarrow$ represent spin-up and spin-down states, and $u$ and $o$ correspond to unoccupied and occupied states, respectively. The denominator $\varepsilon_{u^{\downarrow}} - \varepsilon_{o^{\uparrow/\downarrow}}$ represents the energy separation between the states, while the difference between the matrix elements, $\left| \langle o^{\downarrow} | L_z | u^{\downarrow} \rangle \right|^2 
    - \left| \langle o^{\downarrow} | L_x | u^{\downarrow} \rangle \right|^2$ and $\left| \langle o^{\uparrow} | L_z | u^{\downarrow} \rangle \right|^2 
- \left| \langle o^{\uparrow} | L_x | u^{\downarrow} \rangle \right|^2$, is listed in Table~\ref{matrix} for the Te $p$ orbitals. These quantities allow us to quantify the orbital contributions to the MAE.  

\begin{table}[h!]
   \caption{\justifying Matrix element differences between the two magnetization directions as defined in Eqs.~\ref{oma} and ~\ref{oma2}, i.e., $\left| \langle o^{\downarrow} | L_z | u^{\downarrow} \rangle \right|^2 
    - \left| \langle o^{\downarrow} | L_x | u^{\downarrow} \rangle \right|^2$ and $\left| \langle o^{\uparrow} | L_z | u^{\downarrow} \rangle \right|^2 
- \left| \langle o^{\uparrow} | L_x | u^{\downarrow} \rangle \right|^2$.}
\begin{ruledtabular}
\begin{tabular}{ccccccc}
& \multicolumn{3}{c}{$o^{\downarrow}$} & \multicolumn{3}{c}{$o^{\uparrow}$}  \\ \cline{2-4} \cline{5-7} 
  {$u^{\downarrow}$} & $p_x$ & $p_y$ & $p_z$ & $p_x$ & $p_y$ & $p_z$ \\ \hline
 $p_x$ & 0  & -1  & 0 & 0  & 1 & 0  \\
 $p_y$ & -1  & 0  & 1  & 1 & 0  & -1  \\
 $p_z$ & 0 & 1  & 0  & 0  & -1  & 0  
\label{matrix}
\end{tabular}
\end{ruledtabular}
\end{table}

Equations~\ref{oma} and ~\ref{oma2} highlight that the states closest to the Fermi level dominate the MAE. They further indicate that both intra-spin (same spin channel) and inter-spin (opposite spin channels) excitations between occupied and unoccupied states contribute. To examine the strain and correlation dependence, we analyze the orbital-resolved hybridization contributions to the MAE for strains of $-2\%$, $0\%$, and $+2\%$ under Hubbard $U = 2$, $3$, and $4$ eV, as shown in Fig.~\ref{orb} (a) – (c). The MAE is evaluated using the matrix element differences listed in Table~\ref{matrix} together with the energy denominators from Eqs.~\ref{oma} and \ref{oma2}. In all three cases, the $p_x$–$p_y$ matrix element difference is negative, yielding an IPMA, whereas the $p_y$–$p_z$ matrix difference is positive, favoring OPMA.  

To interpret the MAE variations in Fig.~\ref{orb}, we further examine the orbital projected density of states (PDOS) of Te in the FeTe$_2$ monolayer under the same strain and $U$ conditions, as shown in Fig.~\ref{dos-suppl}. The energy difference between $p{_y^{u^{\downarrow}}}$ and both $p{_z^{o^{\downarrow}}}$ and $p{_z^{o^{\uparrow}}}$ decreases from Fig.~\ref{orb}(a) to Fig.~\ref{orb}(c), as the unoccupied $p_y$ and $p_z$ states shift toward the Fermi level. At $U=4$ eV and $+2\%$ strain [Fig.~\ref{orb}(c)], the $p{_z^{u^{\uparrow}}}$ state crosses the Fermi level, reducing the energy separation $\Delta_1$. Consequently, the $p_y$–$p_z$ matrix element difference increases, thereby enhancing OPMA.  

Conversely, the energy difference between $p{_x^{u^{\downarrow}}}$ and both $p{_y^{o^{\downarrow}}}$ and $p{_y^{o^{\uparrow}}}$ increases with strain and correlation. This occurs because the $p_x^{\downarrow}$ orbital, initially unoccupied at lower $U$, shifts below the Fermi level and becomes occupied, leading to a larger separation $\Delta_2$, from $p_y$ states. As a result, the $p_x$–$p_y$ contribution to IPMA is suppressed, and OPMA becomes dominant at $U=4$ eV and $+2\%$ strain. Notably, for the pristine case ($U=3$ eV, zero strain), the $p_x$–$p_y$ term still dominates over $p_y$–$p_z$, giving rise to an easy-plane anisotropy.  

The giant increase in MAE under these conditions originates from the enhanced out-of-plane orbital mixing facilitated by tensile strain and strong electron correlation, which collectively reshape the crystal field environment and orbital overlaps in the system. This analysis clearly demonstrates that Te orbitals and their hybridization dynamics play a pivotal role in determining the spin orientation in strained FeTe$_2$ monolayers.
 
\subsection{Dzyaloshinskii-Moriya interactions }
\begin{figure}
    \centering
    \includegraphics[width=1.1\linewidth]{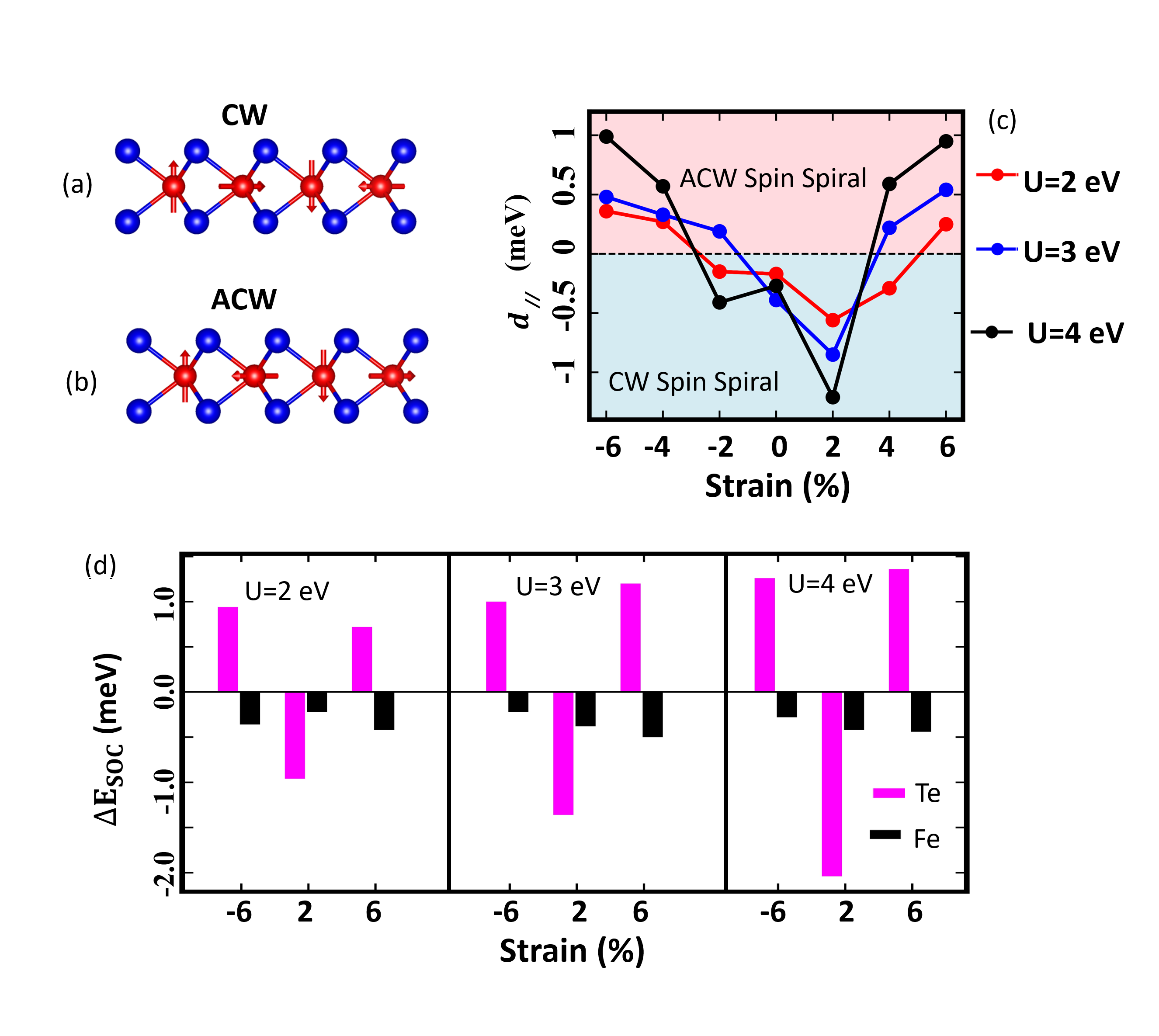}
   \caption{\justifying (a)--(b)  Representation of the Clockwise and Anticlockwise spin configurations in a 4$\times$1 FeTe$_2$ monolayer. Red arrows indicate the directions of the magnetic moments,  
  (c) In-plane DMI component ($d_{\parallel}$) versus strain; positive values favor ACW and negative values favor CW spirals.    
(d) Contribution to $\Delta E_{\text{SOC}}$ by Fe and Te, calculated for opposite chiralities at $\varepsilon = -6\%$, 2\%, and 6\% for $U = 2$, 3, and 4~eV.}
\label{kd}
\end{figure}

\begin{table*}
 \caption{Nearest Neighbor (NN) in-plane and out-of-plane DMI constants
(\(d_{\parallel}\) and \(d_{\perp}\)) and micromagnetic DMI strength
(\(D\)) as a function of strain (\(-6\%\) to \(+6\%\))
calculated using PBE+\(U\) with \(U = 2, 3, 4\) eV.
The units of \(d_{\parallel}\) and \(d_{\perp}\) are meV,
while \(D\) is given in mJ/m\(^2\).}

\begin{ruledtabular}
\begin{tabular}{cccccccccccccccc}
& \multicolumn{3}{c}{U = 2} & \multicolumn{3}{c}{U = 3} & \multicolumn{3}{c}{U = 4} \\ \cline{2-4} \cline{5-7} \cline{8-10}
 Starin ($\%$)& d$_{\parallel}$ & d$_{\perp}$ & D & d$_{\parallel}$ & d$_{\perp}$ & D & d$_{\parallel}$ & d$_{\perp}$ & D\\ \hline
{-6} & 0.36 &0.01&0.99 &0.48 &0.11&1.32&0.99&0.13  &2.72& \\
{-4} & 0.27  &-0.05&0.73&0.33&0.03&0.88&0.57&0.09 &1.53 &\\
{-2} & -0.15 &-0.07&-0.40&0.19&-0.05 &0.50&-0.41&0.02 &-1.08& \\
{0} & -0.17 &-0.08&-0.44&-0.39&-0.04&-1.00&-0.27&-0.08&-0.70& \\
{2} & -0.56 &-0.08&-1.41&-0.85&-0.09&-2.15&-1.21&-0.03&-3.06& \\
{4} & -0.29 &0.05 &-0.72&0.22&0.01&0.54&0.59&0.13&1.46&\\
{6} & 0.25  &0.04&0.61& 0.54 &0.03&1.31&0.95&0.13&2.31&\\
\label{dm-table}
\end{tabular}
\end{ruledtabular}
\end{table*}
 The DMI arises from SOC in systems lacking inversion symmetry, and play a pivotal role in stabilizing non-collinear spin textures (e.g.  skyrmions) which are increasingly being seen as promising for spintronics~\cite{10.1093/nsr/nwy109}. In FeTe$_2$, the stacking of Te layers breaks the mirror symmetry with respect to the Fe atomic plane (horizontal mirror). Following the Moriya's symmetry rules~\cite{dmi,dmi2}, the DMI vector for layer \( n \) can be expressed as 
$$ d_{ij}^{n} = d_{\parallel}^{n}(\hat{z} \times \hat{u}_{ij}) + d_{\perp}^{n}\hat{z},$$  
where \( \hat{z} \) denotes the unit vector along the out-of-plane direction, and \( \hat{u}_{ij} \) is the unit vector connecting sites \( i \) and \( j \). Here, \( d_{\parallel} \) and \( d_{\perp} \) represent the in-plane and out-of-plane components of the DMI strength, respectively. The DMI magnitude can be extracted by computing the energy difference between clockwise (CW) and anticlockwise (ACW) spin spirals (schematic picture of CW and ACW has been shown in Fig. ~\ref{kd} (a)-(b) ) as 
\begin{equation}
d = \frac{E_{\text{CW}} - E_{\text{ACW}}}{m},
\end{equation}
where the prefactor \( m \) depends on the crystal structure and the wavelength of the spin spiral, with \( m = 12 \) for \( d_{\parallel} \) and \( m = 16 \) for \( d_{\perp} \) in our \( 4 \times 1 \) supercell to consider Nearest Neigbour (NN) DMI vector. For details, see section-I of Supplementary material~\cite{SM}. 

The data in Table~\ref{dm-table} show that the out-of-plane DMI component \( d_{\perp} \) remains negligible across all considered strain values and Hubbard \( U \) parameters, indicating that it plays a minimal role in the magnetic interaction landscape of the system. In contrast, the in-plane DMI component \( d_{\parallel} \) exhibits a pronounced strain dependence—first decreasing and then increasing under tensile strain. As shown in Fig.~\ref{kd} (c), \( d_{\parallel} \) changes its sign with strain, switching the preferred spin spiral configuration from ACW to CW, and then back to ACW as shown in Fig. ~\ref{kd} (c). This nonmonotonic behavior arises from strain-induced changes in spin–orbit coupling strength, orbital hybridization, and local crystal symmetry, which together reshape the anisotropic exchange pathways.
The increase in \( d_{\parallel} \) after 2 $\%$ strain, can be attributed to enhanced SOC effects, which originate from the altered electronic hybridization between Fe \( d \)-orbitals and Te \( p \)-orbitals. As strain modifies the d$_{Fe-Te}$, $\theta_{1}$, $\theta_{2}$ within the FeTe\(_2\) monolayer, the resulting change in orbital overlap strengthens the asymmetric exchange interaction responsible for DMI. The inversion asymmetry in the system also increases with strain, further amplifying the DMI strength. For U= 2 eV, at -2 $\%$ strain, \( d_{\parallel} \) is approximately -0.17 meV, but with increasing tensile strain, it rises steadily, reaching 0.25 meV at 6 $\%$ strain, more than  its unstrained value as well as changes the chiral spin from CW to ACW. To verify the nearest-neighbor (NN) approximation for the Dzyaloshinskii--Moriya interaction (DMI) in H-FeTe$_2$, we calculated the second-nearest-neighbor (2NN) in-plane DMI under strains from $-6\%$ to $+6\%$ and Hubbard $U=2$ - 4~eV (Table~S1, Supplementary Material ~\cite{SM}) by taking 4 $\times$ 4 supercell. The 2NN contributions are at least an order of magnitude smaller than the NN DMI, and their inclusion does not affect the strain-driven variation of the DMI or the spin-reorientation transition. This confirms that the DMI in H-FeTe$_2$ is dominated by NN interactions, in contrast to materials such as Fe$_5$GeTe$_2$~\cite{PhysRevB.109.L220404}, where longer-range terms are significant.

To understand the microscopic origin of the DMI, we analyze the atom resolved SOC energy difference  between CW and ACW spin configurations, \( \Delta E_{\text{SOC}} = E_{\text{SOC}}^{\text{ACW}} - E_{\text{SOC}}^{\text{CW}} \),  under selected strain values (\(-6\%\), \(2\%\), and \(6\%\)) for \( U = 2 \), 3, and 4 eV, as shown in Fig.~\ref{kd}(d). 
 The \( \Delta E_{\text{SOC}} \), decomposed into contributions from Fe and Te atoms, reveals a clear trend across all \( U \) values. Te atoms consistently dominate the SOC response, highlighting their central role in generating in-plane DMI. This is primarily due to the larger atomic number of Te, which enhances relativistic effects and thereby amplifies the SOC-driven anisotropic exchange.
Interestingly, the relative signs and magnitudes of Fe and Te contributions vary with strain. At \( -6\% \) strain, Fe and Te contributions have oppositely signed — Fe is positive while Te is negative — suggesting a competing interplay between the two atomic sites that reduces the net DMI and shifts the orientation to ACW.  At \( 2\% \) strain, both Fe and Te contribute positively to \( \Delta E_{\text{SOC}} \), indicating a cooperative enhancement of the CW spin spiral state.  By \( 6\% \) strain, it shows the same behaviour as --6 $\%$ and confirms the ACW spin orientation. For all the U values, the sign of \(\Delta E_{\text{SOC}}\) does not change, however, the strength changes.  The sign reversal of \( \Delta E_{\text{SOC}} \) align with the observed switching behavior of \( d_{\parallel} \) under strain (Fig.~\ref{kd} (d)), further confirming that the evolution of DMI is closely linked to how strain modulates SOC at atomic sites through changes in crystal symmetry, bonding environment, and orbital hybridization.

In the micromagnetic limit with a smooth, spatially varying
magnetization field treated as a continuous vector $\vec{m(r)}$ instead
of a localized spin $\vec{S_i}$, so the DMI energy contribution coming from the third term of the Hamiltonian (Eq. ~\ref{Hamiltomnian}) can be written as:
\begin{equation}
    E_{DMI} = \int dr \epsilon_{DMI} = \int dr \sum_{\mu} D_{\mu}\cdot(\vec{m}\times \partial_{\mu}\vec{m})
\end{equation}
where $\epsilon_{\text{DMI}} = \sum_{\mu} D_{\mu}\cdot(\vec{m}\times \partial_{\mu}\vec{m})$ is the micromagnetic DMI energy density and $\partial_{\mu}$ is the spatial derivative along direction $\mu$ and D$_{\mu}$ is the micromagnetic DMI vector.
For the corresponding micromagnetic
cases, the micromagnetic parameter D$_{\mu}$ is characterized by a single
parameter D with the Interfacial DMI energy reduces to ~\cite{iwasaki2013current,sampaio2013nucleation,fert2013skyrmions,yang2023first}
\begin{equation}
  E_{\text{DMI}} = \int dr \, D \left[ \hat{\mathbf{m}} \cdot (\nabla \times \hat{\mathbf{m}}) - (\hat{\mathbf{m}} \cdot \nabla) \hat{\mathbf{m}} \right] \cdot \hat{\mathbf{z}}.
  \label{dmi}
  \end{equation}
where the micromagnetic DMI coefficient \( D \) is related to the atomic-scale \( d_{\parallel} \) (determined in the Supplementary Material~\cite{SM} (Eq. S19)) by the expression:
\begin{equation}
     D = \frac{\sqrt{3} d_{\parallel} N}{a \cdot t}.
\end{equation}
Here, \( N \) represents the number of magnetic atoms per unit cell, \( a \) is the lattice constant of the hexagonal structure, and \( t \) is the monolayer thickness. 
 As shown in Table~\ref{dm-table}, for each \( U \), \( D \) increases under large compressive (\(-6\%\)) and tensile (\(+6\%\)) strain, reaching its peak magnitude at \( U = 4 \) eV with \( D = 2.72 \) mJ/m$^2$ and \( 2.31 \) mJ/m$^2$, respectively. This indicates that extreme strain conditions amplify the asymmetric exchange interactions responsible for the DMI. In contrast, near-zero strain regimes (especially at \( 2\% \)) exhibit a reversal in the sign of \( D \), reaching highly negative values—up to \( -3.06 \) mJ/m$^2$ at \( U = 4 \) eV — signaling a CW spin spiral orientation. This chirality switching arises from strain induced changes in Fe–Te bond angles and the electronic structure, which alter the strength and direction of SOC contributions. Additionally, increasing \( U \) consistently enhances the magnitude of \( D \), particularly under strain, due to the more localized nature of the Fe \( d \)-orbitals, which strengthens the SOC-mediated exchange interactions. Overall, the micromagnetic DMI constant \( D \) serves as a tunable parameter, sensitively controlled by both strain and electronic correlation, making it a powerful handle for engineering chiral spin textures.
 
 \section{Conclusions}
In summary, we have studied how biaxial strain and electronic correlation affect the magnetic anisotropy energy (MAE) and Dzyaloshinskii–Moriya interaction (DMI) in a pure FeTe$_2$ monolayer using DFT+$U$ with spin–orbit coupling. Our results show that, strain and correlation substantially influence the MAE. Specifically, a tensile strain can switch the magnetic easy axis from in-plane to out-of-plane, with the MAE changing from $\approx$ -6 to + 5.3 meV.  This giant change is driven by shifting of the spin-polarzied occupied and unoccupied stated in the vicinity of the Fermi level. The shifting occurs due to the change in the electron hopping strengths as the bond length changes with strain.  Concurrently, we find that the in-plane DMI component ($d_{\parallel}$), which influences spin twisting, varies strongly with strain. It changes both in sign and strength due to how strain modifies SOC and symmetry. The micromagnetic DMI strength ($D$) reaches up to 3.06 mJ/m$^2$ under certain conditions. In contrast, the out-of-plane component ($d_{\perp}$) stays very small, confirming that the DMI is mainly of the N\'eel-type, driven by in-plane inversion symmetry breaking. Overall, our study shows that both MAE and DMI can be effectively tuned in a pristine 2D material without going through the conventional routes such as chemical doping or tailoring a heterostructure. Therefore, it offers a promising platform to explore complexities of magnetic interactions in 2D systems.
\section{Acknowledgments}
DR Acknowledges the funding from Center for Atomistic Modeling and Material Design to carry out research at IIT Madras. The authors thank Amit Chauhan and Arijit Mandal for useful discussions.
\twocolumngrid
\bibliography{reference.bib}

\end{document}